\documentclass{aa}
\usepackage[varg]{txfonts}
\usepackage{natbib}
\usepackage{url}
\usepackage{hyperref}
\usepackage{xcolor}
\usepackage{ulem}
\usepackage{orcidlink}
\usepackage{xspace}

\usepackage{upgreek}
\usepackage{graphicx}   
\usepackage{multirow}

\newcommand{\src}{{IGR~J06074$+$2205}\xspace}
\newcommand{\nustar}{\textit{NuSTAR}\xspace}
\newcommand{\integral}{\textit{INTEGRAL}\xspace}

\newcommand{\swift}{\textit{Swift}\xspace}

\newcommand{\xmm}{\textit{XMM-Newton}\xspace}

\def\flux{erg\,s$^{-1}$\,cm$^{-2}$\xspace}
\def\lum{erg\,s$^{-1}$\xspace}

\begin{document}

\title{The low-luminosity state of the Be/X-ray pulsar \src}

\titlerunning{Low-luminosity state of \src}

\authorrunning{Salganik, A., et al.}

\author{Alexander~Salganik\inst{\ref{in:UTU}}\orcidlink{0000-0003-2609-8838} \corrauth{alsalganik@gmail.com}, Sergey S. Tsygankov\inst{\ref{in:UTU},\ref{in:KL}}\orcidlink{0000-0002-9679-0793}, Roman Krivonos\inst{\ref{in:IKI}}\orcidlink{0000-0003-2737-5673}, Alexander A. Mushtukov\inst{\ref{in:UCL},\ref{in:OXF}}\orcidlink{0000-0003-2306-419X}, Juri~Poutanen\inst{\ref{in:UTU}}\orcidlink{0000-0002-0983-0049}}
\institute{
Department of Physics and Astronomy, 20014 University of Turku,  Finland 
\label{in:UTU}
\and 
Key Laboratory of Particle Astrophysics, Institute of High Energy Physics, Chinese Academy of Sciences, Beijing 100049, China \label{in:KL}
\and
Space Research Institute (IKI), 84/32 Profsouznaya str., Moscow 117997, Russian Federation\label{in:IKI}
\and
Mullard Space Science Laboratory, University College London, Holmbury St. Mary, Surrey RH5 6NT, UK
\label{in:UCL}
\and
Astrophysics, Department of Physics, University of Oxford, Denys Wilkinson Building, Keble Road, Oxford OX1 3RH, UK
\label{in:OXF}
}
\date{Received 22 September 2026 / Accepted 29 October 2026}

\abstract
{At low accretion rates, X-ray pulsars can develop a two-hump spectrum through the thermalization and Comptonization of cyclotron photons in an accretion-heated, strongly magnetized atmosphere. 
Sources with independently measured cyclotron-line energies, $E_{\rm cyc}$, are particularly important because they provide a direct estimate of magnetic field strength and allow the continuum shape to be tested against spectral formation models.}
{We investigate the luminosity-dependent spectral evolution of the Be/X-ray pulsar \src\ and constrain the accretion regime in its newly observed low-luminosity state.}
{We analyzed three \nustar\ observations obtained during the 2023 and 2025 outbursts, corresponding to the high-, intermediate-, and low-luminosity states at $L_{3-79}\simeq5\times10^{36}$, $4\times10^{35}$, and $6\times10^{34}$~\lum, respectively. 
We supplemented these data with \swift/XRT observations and performed broadband spectral analysis of all three states.}
{The high-luminosity spectrum follows a classical cutoff power law, whereas the intermediate- and low-luminosity spectra require two thermal Comptonization components to reproduce their two-hump shape. 
In the intermediate state, the cyclotron line, first reported at $\sim50$~keV in the high-luminosity  state, lies on top of the high-energy hump. We interpret the low-luminosity state as representing stable accretion from a cold disk. \src\ thus joins the very limited sample of X-ray pulsars with cyclotron-line energies that have been observed with a double-humped spectral shape.}
{}

\keywords{accretion, accretion disks -- magnetic fields -- pulsars: individual: \src\ -- stars: neutron -- X-rays: binaries}

\maketitle
\nolinenumbers

\section{Introduction}
At low accretion rates, the broadband spectra of several X-ray pulsars \citep[XRPs;][]{Mushtukov_Tsygankov_2024_Review} depart from the canonical cutoff power-law continuum and develop two broad humps at a few keV and at several tens of keV
\citep{Tsygankov2019b}. The double-hump spectral shape can be explained by radiation emerging from an accretion-heated, strongly magnetized neutron star (NS) atmosphere. 
\citet{Mushtukov2021} considered a scenario in which a substantial fraction of the accretion energy is released through cyclotron photons generated by collisional excitation of atmospheric electrons. 
These photons are subsequently reprocessed in the overheated atmosphere. 
\citet{SokolovaLapa2021} showed that a similar spectral shape can also arise without this additional source of cyclotron photons through Comptonization of thermal radiation. 
Both of the models rely on similar radiative processes while differing in their assumptions about atmospheric structure and the relative contribution of collisionally generated cyclotron photons.

The intersection energy, $E_{\rm int}$, defined as the energy at which the two spectral components of the double-hump spectrum contribute equally to the flux, has been shown to correlate with the magnetic-field strength, providing an indirect estimate of the latter \citep{Zalot2026}. However, only a few XRPs have a low-luminosity two-hump spectrum and an
independently measured cyclotron-line energy, making new examples particularly valuable.

The X-ray binary \src\ was first discovered in 2003 with the JEM-X instrument on board \integral\ during its first recorded outburst \citep{Chenevez2004}. 
A possible Be-type optical counterpart was subsequently proposed by \citet{Halpern2005}, \citet{Masetti2006}, and  \citet{Tomsick2006}, while \citet{Pooley2004} suggested a radio counterpart. 
Optical spectroscopy classified the counterpart as a B0.5Ve star at a distance of about 4.5~kpc \citep{Reig2010}, but this was later revised to $6.0\pm0.6$~kpc \citep{Fortin2022}. 
Long-term MAXI/GSC monitoring revealed recurrent outbursts with a possible orbital period of $\sim$80~d \citep{Mihara2023,Chhotaray2024}.

The source was identified as an XRP in 2017, when coherent pulsations with a period of 373~s were detected with \xmm\ during quiescence \citep{Reig2018}. Two \nustar\ observations obtained during the October 2023 outburst \citep{Semena2023}, at $L_{3-79}\sim10^{36}$ and $\sim10^{35}$~\lum, were described with a cutoff power-law continuum \citep{Roy2024, Chhotaray2024, Tobrej2024}. The  spectrum in a brighter state showed a cyclotron resonant scattering feature (CRSF) at $\sim50$--55~keV \citep{Roy2024, Chhotaray2024}, while the spectrum in a fainter state  required an additional thermal component with $kT\sim1$~keV, and no CRSF was reported. 
Alternatively, \citet{Tobrej2024} modeled the residuals in the fainter spectrum with a broad absorption feature $\sim$10~keV. 
A later \nustar\ observation during the decline of the September 2025 outburst \citep{Nesci2025} detected the source at $L_{3-30}\sim10^{34}$~\lum, with the spectrum described by an absorbed power law plus a 1.2~keV blackbody \citep{Chhotaray2025}. 

In this work, we report the discovery of a two-hump spectral morphology in the lowest-luminosity \nustar\ observation of \src obtained during the 2025 outburst at $L_{\rm 3-79}\simeq6\times10^{34}$~\lum. A similar structure is also present in the 2023 observation at $L_{\rm 3-79}\simeq4\times10^{35}$~\lum. We examine this spectral transition and constrain the accretion regime in the newly observed low-luminosity state.

\section{Observations and data analysis}
The source was observed with \nustar\ \citep{Harrison2013} on
2023 October 8 (ObsID 90901331002; exposure 42.7~ks), 2023 October 15
(ObsID 90901331004; exposure 40.7~ks), and 2025 October 14
(ObsID 91101339002; exposure 40.1~ks). These observations are hereafter
referred to as NuObs1, NuObs2, and NuObs3, respectively.

The data were reduced using the \textsc{HEASoft} software package
(v6.36) together with CALDB version 20260407. 
Source and background
spectra were generated with the \texttt{nuproducts} task within the standard \texttt{nustardas} pipeline. 
Source events were extracted from circular regions with radii of $70\arcsec$, $60\arcsec$, and $45\arcsec$ for NuObs1, NuObs2, and NuObs3, respectively, while background events were extracted from circular regions with radii of $100\arcsec$, $100\arcsec$, and $70\arcsec$, respectively.

All three observations were affected by stray-light contamination \citep{2017JATIS...3d4003M}. The treatment of the contaminated detector regions and its effect on the adopted energy ranges are described in Appendix~\ref{app:straylight}. The \nustar spectra were analyzed in the 3--79~keV energy range for NuObs1 and NuObs2, while for NuObs3 the range was restricted to 3--40~keV, where the source emission remains above the background level. The \swift/XRT spectra were analyzed in the 0.5--10~keV energy range.

\begin{figure}
    \centering
    \includegraphics[width=\columnwidth]{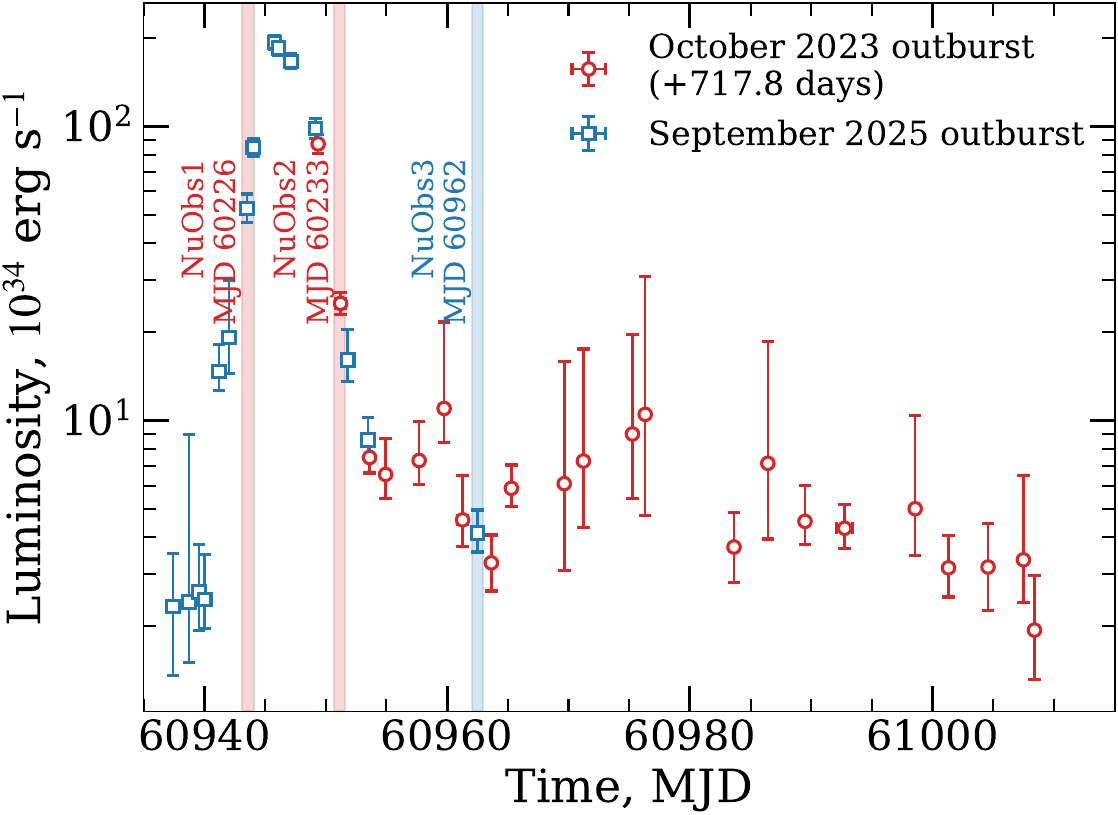}
\caption{\swift/XRT light curves of the outbursts of \src\ in the 0.5--10~keV energy range. The October 2023 outburst has been shifted forward by 717.7 days to align with the September 2025 outburst. The shaded regions indicate the durations of the \nustar observations; NuObs1 and NuObs2 are shown at their corresponding shifted times.}

    \label{fig:overlaid_outbursts}
\end{figure}

\begin{figure}
    \centering
    \includegraphics[width=1.0\linewidth]{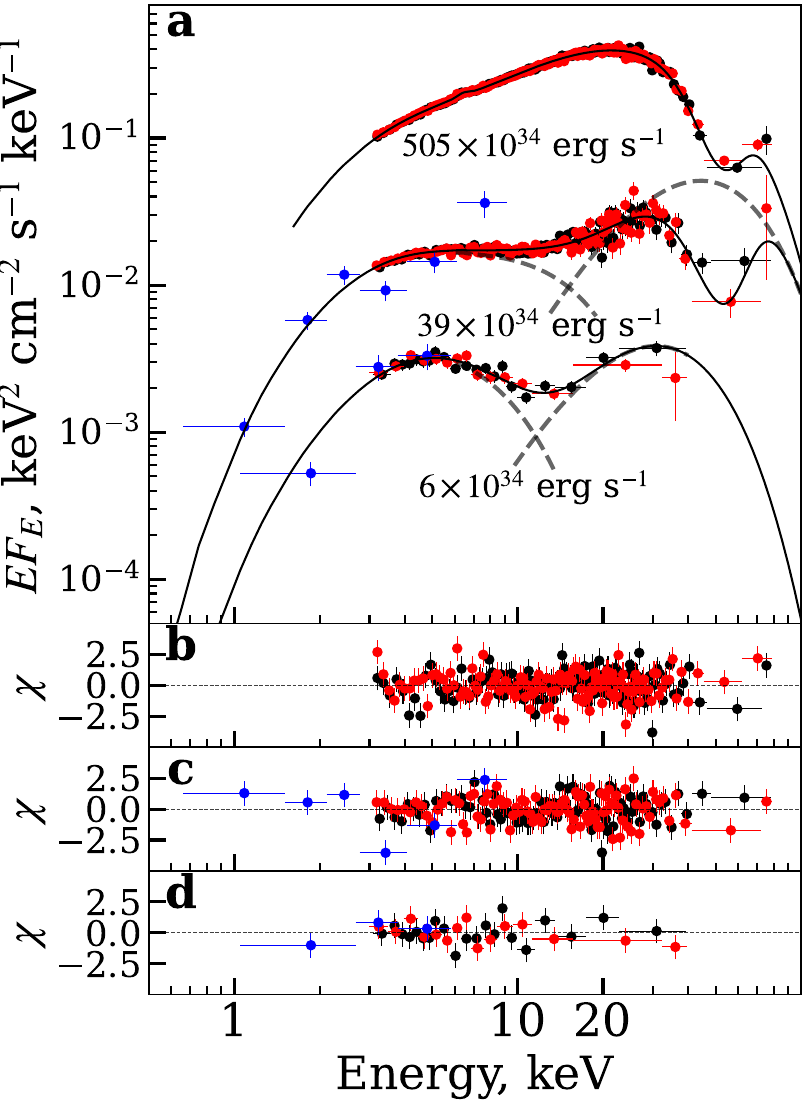}
    \caption{Unfolded broadband \nustar\ and \swift/XRT spectra of \src\ in the low-, intermediate-, and high-luminosity states. The \nustar/FPMA, \nustar/FPMB, and \swift/XRT data are shown in black, red, and blue, respectively. The labels below the spectra indicate the corresponding 3--79~keV luminosities. Solid curves show the best-fit two-component models, while dashed curves show the individual low- and high-energy components. The spectra have also been rebinned for display purposes only. Panels (b--d) show the residuals for the high-, intermediate-, and low-luminosity states, respectively.}    
    \label{fig:sum}
\end{figure}

To trace the October 2023 and September 2025 outbursts, we used
\swift/XRT monitoring data in the 0.5--10~keV band. The 2023 campaign
included 22 observations (ObsIDs 00016281001--011 and
00016281013--023; MJD~60231--60291), while the 2025 campaign
included 15 observations (ObsIDs 00016281052--065 and
03400194001; MJD~60937--60963). The resulting 0.5--10~keV light
curves are shown in Fig.~\ref{fig:overlaid_outbursts}.

For the joint spectral analysis with \nustar, we paired NuObs2 with the closest \swift observation, ObsID~00016281002 at MJD~60233, and NuObs3 with ObsID~03400194001. No simultaneous XRT data were available for NuObs1, and therefore only its \nustar\ spectrum was analyzed.

Throughout this work, all \nustar and \swift spectra were grouped using the optimal binning scheme of \citet{Kaastra2016}, implemented with the \texttt{optmin} option, requiring a minimum of one count per grouped energy bin, and the \texttt{W-statistic} was adopted
for spectral fitting \citep{Wachter1979}. Photoelectric absorption was accounted for using the
abundances of \citet{Wilms2000}. For the two-component
Comptonization fits, the absorption column density was
fixed at $N_{\rm H}=0.4\times10^{22}~\mathrm{cm^{-2}}$,
corresponding to the Galactic value derived from the
HI4PI survey in the direction of the source
\citep{HI4PI2016}. In contrast, $N_{\rm H}$ was left free in the cutoff power-law fits and in the individual \swift/XRT spectral fits used to construct the long-term light curves.

Unless otherwise specified, all uncertainties are given at the 1$\sigma$ confidence level, and all reported fluxes are unabsorbed. Throughout this work, we adopt a source distance of 6~kpc.

\begin{table*}
\centering
\caption{Best-fit spectral parameters for the three luminosity states.}
\label{tab:spectral_param}
\begin{tabular}{lcccccc}
\hline
\hline
Parameter &
\multicolumn{2}{c}{Low state (NuObs3)} &
\multicolumn{2}{c}{Intermediate state (NuObs2)} &
\multicolumn{2}{c}{High state (NuObs1)} \\
&
Low-$E$ comp. & High-$E$ comp. &
Low-$E$ comp. & High-$E$ comp. & Low-$E$ comp. & High-$E$ comp.  \\
\hline
const$_{\rm FPMA}$ &\multicolumn{2}{c}{1.000 (fixed)} & \multicolumn{2}{c}{1.00 (fixed)} & \multicolumn{2}{c}{1.000 (fixed)}\\
const$_{\rm FPMB}$ &\multicolumn{2}{c}{$1.01\pm0.03$} & \multicolumn{2}{c}{$1.01\pm0.01$} & \multicolumn{2}{c}{$1.008\pm0.002$}\\
const$_{\rm XRT}$ & \multicolumn{2}{c}{$0.7\pm0.1$} & \multicolumn{2}{c}{$0.98\pm0.08$} & \multicolumn{2}{c}{--}\\
$kT_{0}$, keV
& \multicolumn{2}{c}{$0.9^{+0.1}_{-0.2}$}
& \multicolumn{2}{c}{$0.66_{-0.07}^{+0.06}$}
&  \multicolumn{2}{c}{$0.78\pm0.01$} \\

$kT_{\rm p}$, keV
& $1.8^{+0.8}_{-0.3}$ & $7.7\pm0.6$
& $2.5^{+0.4}_{-0.2}$ & $7.5_{-0.4}^{+0.8}$
& $4.9^{+0.6}_{-0.3}$ & $9.5^{+0.3}_{-0.4}$ \\

$\tau_{\rm p}$
& $7\pm3$ & 200 (fixed)
& $6.8^{+0.7}_{-0.9}$ & 200 (fixed)
& $6.2^{+0.1}_{-0.2}$ & 200 (fixed)  \\

$E_{\rm K\alpha}$, keV
& \multicolumn{2}{c}{6.4 (fixed)}
& \multicolumn{2}{c}{6.4 (fixed)}
& \multicolumn{2}{c}{6.4 (fixed)} \\

$\sigma_{\rm K\alpha}$, keV
& \multicolumn{2}{c}{0.1 (fixed)}
& \multicolumn{2}{c}{0.1 (fixed)}
& \multicolumn{2}{c}{$0.3\pm0.1$} \\

EW$_{\rm K\alpha}$, keV
& \multicolumn{2}{c}{$<0.1$ (3$\sigma$ up.l)}
& \multicolumn{2}{c}{$<0.04$ (3$\sigma$ up.l)}
& \multicolumn{2}{c}{$0.03\pm0.01$} \\

$E_{\rm cyc}$, keV
& \multicolumn{2}{c}{--}
& \multicolumn{2}{c}{$46.2^{+1.7}_{-1.5}$}
& \multicolumn{2}{c}{$52.1^{+0.5}_{-0.9}$} \\

$\sigma_{\rm cyc}$, keV
& \multicolumn{2}{c}{--}
& \multicolumn{2}{c}{$6^{+2}_{-1}$}
& \multicolumn{2}{c}{$11\pm1$} \\

$\tau_{\rm cyc}$
& \multicolumn{2}{c}{--}
& \multicolumn{2}{c}{$0.8^{+0.6}_{-0.4}$}
& \multicolumn{2}{c}{$1.4^{+0.3}_{-0.4}$} \\

Flux$_{3\text{--}79}$, $10^{-11}$~\flux 
& \multicolumn{2}{c}{$1.35_{-0.08}^{+0.10}$}
& \multicolumn{2}{c}{$9.08\pm0.14$}
& \multicolumn{2}{c}{$117.2\pm0.3$} \\

Luminosity$_{3\text{--}79}$, $10^{34}$~\lum
& \multicolumn{2}{c}{$5.8^{+0.4}_{-0.3}$}
& \multicolumn{2}{c}{$39.1\pm0.6$}
& \multicolumn{2}{c}{$505\pm1$} \\

W-statistic (d.o.f.)
& \multicolumn{2}{c}{211 (195)}
& \multicolumn{2}{c}{421 (384)}
& \multicolumn{2}{c}{462 (437)} \\

\hline
\end{tabular}

\vspace{1mm}
\noindent
\tablefoot{The terms $kT_{\rm p}$ and $\tau_{\rm p}$ denote the plasma temperature and optical depth of the two \texttt{comptt} components. The seed-photon temperature $kT_{0}$ was tied between the low- and high-energy components within each observation. Fluxes and luminosities are unabsorbed and refer to 3--79~keV, assuming a distance of 6 kpc, with its uncertainty neglected. For NuObs3, they were extrapolated from the 3--40~keV fitting range. In all two-component Comptonization fits,
$N_{\rm H}$ was fixed at
$0.4\times10^{22}$~cm$^{-2}$.}
\end{table*}

\section{Results}
\subsection{Long-term light curve}

To investigate the long-term evolution of the source luminosity during the October 2023 and September 2025 outbursts, we analyzed all available \swift/XRT observations obtained during these episodes. 
The XRT spectra were fitted  in the 0.5--10~keV energy range using an absorbed power-law model,
\texttt{tbabs$\times$powerlaw}. The resulting light curves are shown in Fig.~\ref{fig:overlaid_outbursts}. To compare the evolution of the two outbursts directly, the 2023 light curve was shifted in time by 717.7 days to align it with the 2025 event. NuObs1 and NuObs2 respectively sampled the high and intermediate luminosity stages of the 2023 outburst, whereas NuObs3 was obtained near the end of the rapid decline of the 2025 outburst. The similarity between the two light curves suggests that NuObs3 sampled the transition to a low-luminosity state at $L_{\rm 0.5-10} \simeq 4 \times 10^{34}$ erg s$^{-1}$, which is investigated in this work.

\subsection{Spectral analysis} 
As a first step, we attempted to describe all three spectra using a power law with a high-energy cutoff (\texttt{powerlaw$\times$highecut} in \textsc{xspec}), a model commonly applied to accreting XRPs. Because the \texttt{highecut} model is discontinuous at $E_{\rm cut}$, we smoothed the resulting artificial residuals following \citet{Coburn2002} by adding a \texttt{gabs} component with its centroid tied to $E_{\rm cut}$ and width fixed at $\sigma_{\rm gabs}=0.1E_{\rm cut}$. The optical depth was left free to vary. An additional Gaussian component was included in all fits
to account for Fe K$\alpha$ emission near 6.4~keV,
while a second \texttt{gabs} component was included for
NuObs1 and NuObs2 to model the cyclotron absorption feature. To account for cross-calibration uncertainties between FPMA, FPMB, and, where available, \swift/XRT, we also included a multiplicative normalization constant. Thus, the model can be written as
\texttt{const$\times$tbabs$\times$
(powerlaw$\times$highecut$\times$gabs$_{\rm smooth}$+gaussian)$\times$gabs$_{\rm cyc}$},
with the cyclotron component omitted for NuObs3.

Figure~\ref{fig:sum} shows the unfolded broadband spectra of \src\ 
obtained with \nustar\ at three different luminosity levels. The brightest spectrum (NuObs1; $L_{3-79}\simeq5\times10^{36}$~\lum) is adequately described by this baseline model, yielding a W-statistic (d.o.f.) of 455 (437). In contrast, the same model does not provide an acceptable description of the intermediate- and low-luminosity spectra, NuObs2 and NuObs3, for which the W-statistic (d.o.f.) is 628 (383) and 296 (194), respectively. In both cases, the fits show systematic residuals at energies above $\sim$20--40~keV, where the model underestimates the observed flux. 

The intermediate- (NuObs2; $L_{3-79}\simeq4\times10^{35}$~\lum) and low-luminosity (NuObs3; $L_{3-79}\simeq6\times10^{34}$~\lum) spectra instead revealed a two-hump structure. Such a morphology is inconsistent with a single-component cutoff power-law continuum and is commonly observed in accreting XRPs at luminosities of $L_{\rm 3-79}\sim10^{34}$--$10^{36}$~\lum. We therefore modeled the continua of NuObs2 and NuObs3 using a sum of two thermal Comptonization components (\texttt{comptt+comptt}; \citealt{Titarchuk1994}). The seed-photon temperature $kT_0$ was tied between the two components, whereas their electron temperatures were fitted independently. The optical depth of the low-energy component was left free, while that of the high-energy component was fixed at $\tau_{\rm p}=200$ to reduce degeneracy with its electron temperature. This choice is purely phenomenological and does not imply a physically meaningful optical depth. This model reproduces both spectral peaks and provides a statistically acceptable description of both datasets, yielding W-statistic (d.o.f.) values of 421 (384) and 211 (195) for NuObs2 and NuObs3, respectively. The iron K$\alpha$ emission line is not detected in either observation. For NuObs2, the model additionally included a cyclotron absorption component (\texttt{gabs}). Notably, in NuObs2, the cyclotron line lies directly on top of the high-energy hump. No CRSF is reported for NuObs3, as the expected cyclotron-line energy lies outside the adopted 3--40~keV fitting range. 

For consistency, NuObs1 was also fitted with the same phenomenological two-component thermal Comptonization continuum adopted for NuObs2 and NuObs3. The final spectral model was \texttt{const$\times$tbabs$\times$(comptt+comptt+gauss)$\times$gabs}, where the Gaussian component accounts for Fe K$\alpha$ emission and \texttt{gabs} describes the cyclotron absorption feature. The latter was included only for NuObs1 and NuObs2, while the iron line component was included in all three fits. The resulting best-fit spectral parameters are summarized in Table~\ref{tab:spectral_param}.

\section{Discussion}
\label{sec:discussion}

\subsection{Relation between the hump-intersection and cyclotron-line energies}

We analyzed three \nustar\ observations of \src\ spanning nearly two orders of magnitude in luminosity. The spectrum at the highest luminosity, $L_{3-79}\simeq5\times10^{36}$~\lum, is well described by a power law with a high-energy cutoff together with an Fe K$\alpha$ emission line and a CRSF at about 52~keV. At lower luminosities, $L_{3-79}\simeq6\times10^{34}$ and $4\times10^{35}$~\lum, the spectra show a substantially different shape characterized by two broad components. In the intermediate-luminosity state, the CRSF is superimposed on the high-energy hump, making its inferred parameters potentially sensitive to the adopted description of the continuum. More generally, the strong luminosity-dependent evolution of the continuum illustrated in Fig.~\ref{fig:sum} should therefore be kept in mind when CRSF parameters measured in different luminosity states are compared.

A similar spectral configuration was observed in A~0535$+$262, where the CRSF is superimposed on the high-energy hump in the intermediate-luminosity state \citep{Tsygankov2019a,Mushtukov2021}. 
Such a two-hump spectrum is expected to form at low accretion rates, when the accretion flow is decelerated by collisions in the strongly magnetized NS atmosphere. 
The deposited energy heats the upper atmospheric layers, where Compton scattering modifies the emerging radiation. 
In the model of \citet{Mushtukov2021}, cyclotron photons generated during the braking of the accretion flow contribute significantly to the high-energy component, while part of the deposited energy is thermalized and produces the lower-energy component. 
\citet{SokolovaLapa2021} demonstrated that a similar spectral structure can also arise from the Comptonization of thermal photons without an additional source of cyclotron radiation.
Both models involve similar radiative processes, and the relative contribution of different photon sources depends on the atmospheric conditions. 
Our observations did not allow us to distinguish between these scenarios since the phenomenological spectral model used here does not constrain the physical origin of the two components. 
A quantitative comparison would require fitting the data with physical atmosphere models.

In both models, resonant scattering affects the high-energy component, linking its spectral shape to the magnetic field strength and the cyclotron energy.
\citet{Zalot2026} analyzed the low-luminosity spectra of nine Be/X-ray pulsars using an empirical double-hump model consisting of two cutoff power-law components. For the four sources with secure CRSF measurements obtained during outbursts, they compared the cyclotron-line energy with the intersection energy, $E_{\rm int}$, defined as the energy at which the fluxes of the two continuum components are equal. From the weighted average of the four $E_{\rm cyc}/E_{\rm int}$ ratios, they obtained
\begin{equation}
E_{\rm cyc}=(4.15\pm0.11)\,E_{\rm int}.
\end{equation}

To ensure a consistent comparison, we fitted the low-luminosity spectrum using the empirical double-hump model of \citet{Zalot2026} and obtained
$E_{\rm int}=11.7^{+0.6}_{-0.5}$~keV. Employing the cyclotron-line energy measured from the bright-state spectrum,
$E_{\rm cyc}=52.1^{+0.5}_{-0.9}$~keV, we obtained
$E_{\rm cyc}/E_{\rm int}=4.5\pm0.2$. This value is consistent with the relation reported by \citet{Zalot2026}.

We note that \src\ is well suited for this comparison because its two-hump spectrum is observed at a luminosity below $10^{35}$~\lum, that is, within the low-luminosity regime considered by \citet{Zalot2026}. Therefore, the agreement obtained here supports the proposed connection between the characteristic energy of the two-hump spectrum and the magnetic-field strength. 

\subsection{Nature of the low-luminosity state}

The detection of coherent pulsations at low luminosities, both during quiescence \citep{Reig2018} and in the observation at $L_{3-79}\simeq6\times10^{34}$~\lum\ \citep{Chhotaray2025}, together with the two-hump spectrum obtained in this work indicate that matter continues to reach the magnetic poles of the NS. This implies that the source has not entered the propeller regime, in which the rapidly rotating magnetosphere acts as a centrifugal barrier and inhibits accretion onto the NS surface \citep{Illarionov1975,Campana2002}.

A possible interpretation is stable accretion from a cold disk. In this regime, the disk temperature falls below the hydrogen-ionization threshold, while accretion continues, provided that the magnetospheric radius remains smaller than the corotation radius \citep{Tsygankov2017}. Using the cyclotron-line energy, the magnetic-field strength can be estimated as
\begin{equation}
B_{12}
=
\frac{1+z}{11.6}\,E_{\rm cyc}
=
\frac{1}{\sqrt{1-\dfrac{2GM_{\rm NS}}{R_{\rm NS}c^2}}}
\frac{E_{\rm cyc}}{11.6},
\end{equation}
where $B_{12}=B/(10^{12}~{\rm G})$ and $z$ is the gravitational redshift at the NS surface.
For $E_{\rm cyc}=52.1$~keV and canonical NS parameters
$M_{\rm NS}=1.4\,M_{\odot}$ and $R_{\rm NS}=12$~km, this gives
$B_{12}\simeq6$. With $P_{\rm spin}\simeq375$~s, \src\ lies in the cold-disk accretion region of the magnetic-field--spin-period diagram in Fig.~3 of \citet{Tsygankov2017}. Its long spin period therefore allows the disk to become cold before the onset of the propeller regime.

We estimated the luminosity corresponding to the transition to a cold disk using the refined criterion derived by \citet{Tsygankov2017}:
\begin{equation}
L_{\rm cold}\simeq
7\times10^{33}
A^{-7/13}k^{21/13}M_{1.4}^{3/13}R_6^{23/13}
B_{12}^{12/13}T_{6500}^{28/13}\ {\rm erg\ s^{-1}},
\end{equation}
where $M_{1.4}=M_{\rm NS}/(1.4\,M_{\odot})$, $R_6=R_{\rm NS}/(10^6\,{\rm cm})$, and $T_{6500}=T_{\rm eff}/6500\,{\rm K}$. The parameter $k=R_{\rm m}/R_{\rm A}$ describes the disk--magnetosphere coupling, and we adopted the commonly used value $k=0.5$ for disk accretion \citep{Ghosh1978}, while $A=0.057$ corresponds to the case where the viscous stress vanishes at the magnetospheric radius. Adopting canonical values for the NS mass and radius and $B_{12}\simeq6$ as inferred from the CRSF energy, we obtained
$L_{\rm cold}\simeq8\times10^{34}$~\lum.

We stress that the value obtained above represents the characteristic transition luminosity at which the accretion disk is expected to enter the cold state rather than the luminosity of the source after the transition has already occurred. This value is close to the luminosity measured during the lowest-luminosity observation,
$L_{3-79}\simeq6\times10^{34}$~\lum. We also note that the comparison is necessarily approximate since the theoretical transition luminosity is bolometric and depends on the adopted values of the disk--magnetosphere coupling parameter and on the assumptions regarding the inner boundary of the accretion disk. Nevertheless, considering these uncertainties, the measured luminosity is in broad agreement with the expected transition luminosity.

\section{Conclusions}
\label{sec:conclusions}

We analyzed three \nustar\ observations of \src\ spanning nearly two orders of magnitude in X-ray luminosity. Our main results are summarized as follows:

\begin{itemize}

\item The high-luminosity spectrum at
$L_{3\text{--}79}\simeq5\times10^{36}$~\lum\ is adequately described by the canonical cutoff power-law continuum commonly used for accreting XRPs. In contrast, the same single-component model does not reproduce the intermediate- and low-luminosity spectra observed at
$L_{3\text{--}79}\simeq4\times10^{35}$ and
$6\times10^{34}$~\lum, respectively. These spectra show a two-hump structure, with the CRSF superimposed on the high-energy hump in the intermediate state.

\item We interpret this spectral evolution as a transition to the low-accretion regime in which radiation pressure is insufficient to decelerate the accretion flow above the NS surface. In this regime, the spectrum is formed in the strongly magnetized NS atmosphere, with cyclotron-related processes contributing to the high-energy component.

\item For the low-luminosity spectrum, we obtained an intersection energy of
$E_{\rm int}=11.7^{+0.6}_{-0.5}$~keV. Together with the cyclotron-line energy measured in the bright state,
$E_{\rm cyc}=52.1^{+0.5}_{-0.9}$~keV,  this gives
$E_{\rm cyc}/E_{\rm int}=4.5\pm0.2$, which is consistent with the relation reported by \citet{Zalot2026}.

\item The detection of coherent pulsations and the two-hump spectrum at
$L_{3\text{--}79}\simeq6\times10^{34}$~\lum\ indicate that accretion onto the NS surface continues and that the source has not entered the propeller regime. The observed luminosity is close to the estimated cold-disk transition luminosity,
$L_{\rm cold}\simeq8\times10^{34}$~\lum, supporting the interpretation of the low-luminosity state as stable accretion from a cold disk.

\end{itemize}

\begin{acknowledgements}
This research made use of the \nustar Data Analysis Software (NUSTARDAS) jointly developed by the ASI Science Data Centre (ASDC, Italy) and Caltech. This work also made use of data supplied by the UK \swift Science Data Centre at the
University of Leicester.
AS acknowledges support from the Jenny and Antti Wihuri Foundation (grant no. 00250351).
AAM acknowledges support from the UKRI1169 Astronomy observation and theory small award.
This research was supported by the International Space Science Institute (ISSI) in Bern, through International Team project 25-657 `Polarimetric Insights into Extreme Magnetism', the  Research Council of Finland Centre of Excellence in Neutron-Star Physics (grant 374064), and the grant of Ministry of Science and Higher Education of RF  075-15-2024-647.
\end{acknowledgements}

\bibliography{allbib}

@ARTICLE{Chenevez2004,
       author = {{Chenevez}, J. and {Budtz-Jorgensen}, C. and {Lund}, N. and {Westergaard}, N.~J. and {Kretschmar}, P. and {Rodriguez}, J. and {Orr}, A. and {Hermsen}, Wim},
        title = "{IGR J06074+2205 a new X-ray source discovered by INTEGRAL}",
      journal = {The Astronomer's Telegram},
         year = 2004,
        month = jan,
       volume = {223},
        pages = {1},
       adsurl = {https://ui.adsabs.harvard.edu/abs/2004ATel..223....1C}
}

@ARTICLE{Tomsick2006,
       author = {{Tomsick}, J.~A. and {Chaty}, S. and {Rodriguez}, J. and {Walter}, R. and {Kaaret}, P.},
        title = "{IGR J06074+2205 is a Be X-ray Binary}",
      journal = {The Astronomer's Telegram},
         year = 2006,
        month = dec,
       volume = {959},
        pages = {1},
       adsurl = {https://ui.adsabs.harvard.edu/abs/2006ATel..959....1T}
}

@ARTICLE{Halpern2005,
       author = {{Halpern}, J.~P. and {Tyagi}, S.},
        title = "{Be Star Counterpart of X-ray Transient IGR J06074+2205}",
      journal = {The Astronomer's Telegram},
         year = 2005,
        month = dec,
       volume = {682},
        pages = {1},
       adsurl = {https://ui.adsabs.harvard.edu/abs/2005ATel..682....1H}
}

@ARTICLE{Masetti2006,
       author = {{Masetti}, N. and {Bassani}, L. and {Bazzano}, A. and {Bird}, A.~J. and {Dean}, A.~J. and {Malizia}, A. and {Norci}, L. and {Palazzi}, E. and {Schwope}, A.~D. and {Stephen}, J.~B. and {Ubertini}, P. and {Walter}, R.},
        title = "{Unveiling the nature of INTEGRAL objects through optical spectroscopy. IV. A study of six new hard X-ray sources}",
      journal = {\aap},
         year = 2006,
        month = aug,
       volume = {455},
       number = {1},
        pages = {11-19},
          doi = {10.1051/0004-6361:20065111},
archivePrefix = {arXiv},
       eprint = {astro-ph/0604482},
 primaryClass = {astro-ph},
       adsurl = {https://ui.adsabs.harvard.edu/abs/2006A&A...455...11M}
}

@ARTICLE{Pooley2004,
       author = {{Pooley}, Guy},
        title = "{IGR J06074+2205 - possible association with radio source}",
      journal = {The Astronomer's Telegram},
         year = 2004,
        month = jan,
       volume = {226},
        pages = {1},
       adsurl = {https://ui.adsabs.harvard.edu/abs/2004ATel..226....1P}
}

@ARTICLE{Reig2010,
       author = {{Reig}, P. and {Zezas}, A. and {Gkouvelis}, L.},
        title = "{The optical counterpart to IGR J06074+2205: a Be/X-ray binary showing disc loss and V/R variability}",
      journal = {\aap},
         year = 2010,
        month = nov,
       volume = {522},
          eid = {A107},
        pages = {A107},
          doi = {10.1051/0004-6361/201014788},
archivePrefix = {arXiv},
       eprint = {1006.4935},
 primaryClass = {astro-ph.GA},
       adsurl = {https://ui.adsabs.harvard.edu/abs/2010A&A...522A.107R}
}

@ARTICLE{Fortin2022,
       author = {{Fortin}, Francis and {Garc{\'\i}a}, Federico and {Chaty}, Sylvain},
        title = "{Finding the birthplace of HMXBs in the Galaxy using Gaia EDR3: Kinematical age determination through orbit integration}",
      journal = {\aap},
         year = 2022,
        month = sep,
       volume = {665},
          eid = {A69},
        pages = {A69},
          doi = {10.1051/0004-6361/202244048},
archivePrefix = {arXiv},
       eprint = {2207.02114},
 primaryClass = {astro-ph.GA},
       adsurl = {https://ui.adsabs.harvard.edu/abs/2022A&A...665A..69F}
}

@ARTICLE{Reig2018,
       author = {{Reig}, P. and {Zezas}, A.},
        title = "{Discovery of X-ray pulsations in the Be/X-ray binary IGR J06074+2205}",
      journal = {\aap},
         year = 2018,
        month = may,
       volume = {613},
          eid = {A52},
        pages = {A52},
          doi = {10.1051/0004-6361/201732533},
archivePrefix = {arXiv},
       eprint = {1802.04552},
 primaryClass = {astro-ph.HE},
       adsurl = {https://ui.adsabs.harvard.edu/abs/2018A&A...613A..52R}
}

@ARTICLE{Roy2024,
       author = {{Roy}, Kinjal and {Sharma}, Rahul and {Manikantan}, Hemanth and {Paul}, Biswajit},
        title = "{Detection of a cyclotron line in the Be X-ray pulsar IGR J06074+2205}",
      journal = {\aap},
         year = 2024,
        month = jun,
       volume = {686},
          eid = {A145},
        pages = {A145},
          doi = {10.1051/0004-6361/202348998},
archivePrefix = {arXiv},
       eprint = {2402.17382},
 primaryClass = {astro-ph.HE},
       adsurl = {https://ui.adsabs.harvard.edu/abs/2024A&A...686A.145R}
}

@ARTICLE{Tobrej2024,
       author = {{Tobrej}, Mohammed and {Rai}, Binay and {Ghising}, Manoj and {Paul}, Bikash Chandra},
        title = "{NuSTAR detection of a broad absorption line in IGR J06074+2205}",
      journal = {JHEA},
         year = 2024,
        month = jun,
       volume = {42},
        pages = {129-135},
          doi = {10.1016/j.jheap.2024.04.006},
archivePrefix = {arXiv},
       eprint = {2404.14805},
 primaryClass = {astro-ph.HE},
       adsurl = {https://ui.adsabs.harvard.edu/abs/2024JHEAp..42..129T}
}

@ARTICLE{Mushtukov2021,
       author = {{Mushtukov}, Alexander A. and {Suleimanov}, Valery F. and {Tsygankov}, Sergey S. and {Portegies Zwart}, Simon},
        title = "{Spectrum formation in X-ray pulsars at very low mass accretion rate: Monte Carlo approach}",
      journal = {\mnras},
         year = 2021,
        month = may,
       volume = {503},
       number = {4},
        pages = {5193-5203},
          doi = {10.1093/mnras/stab811},
archivePrefix = {arXiv},
       eprint = {2006.13596},
 primaryClass = {astro-ph.HE},
       adsurl = {https://ui.adsabs.harvard.edu/abs/2021MNRAS.503.5193M}
}

@ARTICLE{SokolovaLapa2021,
       author = {{Sokolova-Lapa}, E. and {Gornostaev}, M. and {Wilms}, J. and {Ballhausen}, R. and {Falkner}, S. and {Postnov}, K. and {Thalhammer}, P. and {F{\"u}rst}, F. and {Garc{\'\i}a}, J.~A. and {Shakura}, N. and {Becker}, P.~A. and {Wolff}, M.~T. and {Pottschmidt}, K. and {H{\"a}rer}, L. and {Malacaria}, C.},
        title = "{X-ray emission from magnetized neutron star atmospheres at low mass-accretion rates. I. Phase-averaged spectrum}",
      journal = {\aap},
         year = 2021,
        month = jul,
       volume = {651},
          eid = {A12},
        pages = {A12},
          doi = {10.1051/0004-6361/202040228},
archivePrefix = {arXiv},
       eprint = {2104.06802},
 primaryClass = {astro-ph.HE},
       adsurl = {https://ui.adsabs.harvard.edu/abs/2021A&A...651A..12S}
}

@ARTICLE{Chhotaray2025,
       author = {{Chhotaray}, Birendra and {Naik}, Sachindra and {Jaisawal}, Gaurava K. and {Jana}, Arghajit and {Mandal}, Manoj},
        title = "{NuSTAR observation of the Be/X-ray binary IGR J06074+2205}",
      journal = {The Astronomer's Telegram},
         year = 2025,
        month = oct,
       volume = {17458},
        pages = {1},
       adsurl = {https://ui.adsabs.harvard.edu/abs/2025ATel17458....1C}
}

@ARTICLE{Tsygankov2019a,
       author = {{Tsygankov}, Sergey S. and {Doroshenko}, Victor and {Mushtukov}, Alexander A. and {Suleimanov}, Valery F. and {Lutovinov}, Alexander A. and {Poutanen}, Juri},
        title = "{Cyclotron emission, absorption, and the two faces of X-ray pulsar A 0535+262}",
      journal = {\mnras},
         year = 2019,
        month = jul,
       volume = {487},
       number = {1},
        pages = {L30-L34},
          doi = {10.1093/mnrasl/slz079},
archivePrefix = {arXiv},
       eprint = {1905.09496},
 primaryClass = {astro-ph.HE},
       adsurl = {https://ui.adsabs.harvard.edu/abs/2019MNRAS.487L..30T}
}

@ARTICLE{Nesci2025,
       author = {{Nesci}, Roberto and {Fiocchi}, Mariateresa and {Vagnozzi}, Antonio},
        title = "{Strong X-ray outburst of IGR J06074+2205}",
      journal = {The Astronomer's Telegram},
         year = 2025,
        month = sep,
       volume = {17424},
        pages = {1},
       adsurl = {https://ui.adsabs.harvard.edu/abs/2025ATel17424....1N}
}

@ARTICLE{Chhotaray2024,
       author = {{Chhotaray}, Birendra and {Naik}, Sachindra and {Jaisawal}, Gaurava K. and {Ahuja}, Goldy},
        title = "{Optical and X-ray studies of the Be/X-ray binary IGR J06074+2205}",
      journal = {\mnras},
         year = 2024,
        month = nov,
       volume = {534},
       number = {3},
        pages = {2830-2847},
          doi = {10.1093/mnras/stae2282},
archivePrefix = {arXiv},
       eprint = {2410.00747},
 primaryClass = {astro-ph.HE},
       adsurl = {https://ui.adsabs.harvard.edu/abs/2024MNRAS.534.2830C}
}

@ARTICLE{Mihara2023,
       author = {{Mihara}, T. and {Nakajima}, M. and {Negoro}, H. and {Kobayashi}, K. and {Tanaka}, M. and {Soejima}, Y. and {Kudo}, Y. and {Kawamuro}, T. and {Yamada}, S. and {Tamagawa}, T. and {Kawai}, N. and {Matsuoka}, M. and {Sakamoto}, T. and {Serino}, M. and {Sugita}, S. and {Hiramatsu}, H. and {Nishikawa}, H. and {Yoshida}, A. and {Tsuboi}, Y. and {Urabe}, S. and {Nawa}, S. and {Nemoto}, N. and {Goto}, E. and {Shidatsu}, M. and {Takahashi}, I. and {Niwano}, M. and {Sato}, S. and {Higuchi}, N. and {Yatsu}, Y. and {Nakahira}, S. and {Ueno}, S. and {Tomida}, H. and {Ishikawa}, M. and {Ogawa}, S. and {Kurihara}, T. and {Ueda}, Y. and {Setoguchi}, K. and {Yoshitake}, T. and {Nakatani}, Y. and {Okada}, Y. and {Yamauchi}, M. and {Hagiwara}, Y. and {Umeki}, Y. and {Otsuki}, Y. and {Yamaoka}, K. and {Kawakubo}, Y. and {Sugizaki}, M. and {Iwakiri}, W.},
        title = "{MAXI/GSC observations of weak X-ray outbursts from the Be/X-ray binary pulsar IGR J06074+2205 and the possible binary orbital period}",
      journal = {The Astronomer's Telegram},
         year = 2023,
        month = nov,
       volume = {16351},
        pages = {1},
       adsurl = {https://ui.adsabs.harvard.edu/abs/2023ATel16351....1M}
}

@ARTICLE{Semena2023,
       author = {{Semena}, A.~N. and {Mereminskiy}, I.~A. and {Lutovinov}, A.~A. and {Molkov}, S.~V. and {Arefiev}, V.~A. and {Shtykovsky}, A.~E. and {Tkachenko}, A. Yu. and {Kuznetsova}, E.~A.},
        title = "{ART-XC detection of BeXRB IGR J06074+2205 in outburst}",
      journal = {The Astronomer's Telegram},
         year = 2023,
        month = oct,
       volume = {16267},
        pages = {1},
       adsurl = {https://ui.adsabs.harvard.edu/abs/2023ATel16267....1S}
}

@ARTICLE{Wachter1979,
       author = {{Wachter}, K. and {Leach}, R. and {Kellogg}, E.},
        title = "{Parameter estimation in X-ray astronomy using maximum likelihood.}",
      journal = {\apj},
         year = 1979,
        month = may,
       volume = {230},
        pages = {274-287},
          doi = {10.1086/157084},
       adsurl = {https://ui.adsabs.harvard.edu/abs/1979ApJ...230..274W}
}

@ARTICLE{HI4PI2016,
       author = {{HI4PI Collaboration} and {Ben Bekhti}, N. and {Fl{\"o}er}, L. and {Keller}, R. and {Kerp}, J. and {Lenz}, D. and {Winkel}, B. and {Bailin}, J. and {Calabretta}, M.~R. and {Dedes}, L. and {Ford}, H.~A. and {Gibson}, B.~K. and {Haud}, U. and {Janowiecki}, S. and {Kalberla}, P.~M.~W. and {Lockman}, F.~J. and {McClure-Griffiths}, N.~M. and {Murphy}, T. and {Nakanishi}, H. and {Pisano}, D.~J. and {Staveley-Smith}, L.},
        title = "{HI4PI: A full-sky H I survey based on EBHIS and GASS}",
      journal = {\aap},
         year = 2016,
        month = oct,
       volume = {594},
          eid = {A116},
        pages = {A116},
          doi = {10.1051/0004-6361/201629178},
archivePrefix = {arXiv},
       eprint = {1610.06175},
 primaryClass = {astro-ph.GA},
       adsurl = {https://ui.adsabs.harvard.edu/abs/2016A&A...594A.116H}
}

@ARTICLE{Wilms2000,
       author = {{Wilms}, J. and {Allen}, A. and {McCray}, R.},
        title = "{On the Absorption of X-Rays in the Interstellar Medium}",
      journal = {\apj},
         year = 2000,
        month = oct,
       volume = {542},
       number = {2},
        pages = {914-924},
          doi = {10.1086/317016},
archivePrefix = {arXiv},
       eprint = {astro-ph/0008425},
 primaryClass = {astro-ph},
       adsurl = {https://ui.adsabs.harvard.edu/abs/2000ApJ...542..914W}
}

@ARTICLE{Titarchuk1994,
       author = {{Titarchuk}, Lev},
        title = "{Generalized Comptonization Models and Application to the Recent High-Energy Observations}",
      journal = {\apj},
         year = 1994,
        month = oct,
       volume = {434},
        pages = {570},
          doi = {10.1086/174760},
       adsurl = {https://ui.adsabs.harvard.edu/abs/1994ApJ...434..570T}
}

@ARTICLE{Tsygankov2019b,
       author = {{Tsygankov}, Sergey S. and {Rouco Escorial}, Alicia and {Suleimanov}, Valery F. and {Mushtukov}, Alexander A. and {Doroshenko}, Victor and {Lutovinov}, Alexander A. and {Wijnands}, Rudy and {Poutanen}, Juri},
        title = "{Dramatic spectral transition of X-ray pulsar GX 304-1 in low luminous state}",
      journal = {\mnras},
         year = 2019,
        month = feb,
       volume = {483},
       number = {1},
        pages = {L144-L148},
          doi = {10.1093/mnrasl/sly236},
archivePrefix = {arXiv},
       eprint = {1810.13307},
 primaryClass = {astro-ph.HE},
       adsurl = {https://ui.adsabs.harvard.edu/abs/2019MNRAS.483L.144T}
}

@ARTICLE{Zalot2026,
       author = {{Zalot}, Nicolas and {Sokolova-Lapa}, Ekaterina and {Zainab}, Aafia and {Thalhammer}, Philipp and {Stierhof}, Jakob and {Berger}, Katrin and {Pottschmidt}, Katja and {Ballhausen}, Ralf and {Malacaria}, Christian and {Gulbahar}, Esin and {Wilms}, J{\"o}rn},
        title = "{A simple relation: Neutron star magnetic field strength and spectral shape at low mass accretion rates}",
      journal = {\aap},
         year = 2026,
        month = may,
       volume = {709},
          eid = {A182},
        pages = {A182},
          doi = {10.1051/0004-6361/202558692},
archivePrefix = {arXiv},
       eprint = {2604.14968},
 primaryClass = {astro-ph.HE},
       adsurl = {https://ui.adsabs.harvard.edu/abs/2026A&A...709A.182Z}
}

@ARTICLE{Coburn2002,
       author = {{Coburn}, W. and {Heindl}, W.~A. and {Rothschild}, R.~E. and {Gruber}, D.~E. and {Kreykenbohm}, I. and {Wilms}, J. and {Kretschmar}, P. and {Staubert}, R.},
        title = "{Magnetic Fields of Accreting X-Ray Pulsars with the Rossi X-Ray Timing Explorer}",
      journal = {\apj},
         year = 2002,
        month = nov,
       volume = {580},
       number = {1},
        pages = {394-412},
          doi = {10.1086/343033},
archivePrefix = {arXiv},
       eprint = {astro-ph/0207325},
 primaryClass = {astro-ph},
       adsurl = {https://ui.adsabs.harvard.edu/abs/2002ApJ...580..394C}
}

@ARTICLE{Tsygankov2017,
       author = {{Tsygankov}, S.~S. and {Mushtukov}, A.~A. and {Suleimanov}, V.~F. and {Doroshenko}, V. and {Abolmasov}, P.~K. and {Lutovinov}, A.~A. and {Poutanen}, J.},
        title = "{Stable accretion from a cold disc in highly magnetized neutron stars}",
      journal = {\aap},
         year = 2017,
        month = nov,
       volume = {608},
          eid = {A17},
        pages = {A17},
          doi = {10.1051/0004-6361/201630248},
archivePrefix = {arXiv},
       eprint = {1703.04528},
 primaryClass = {astro-ph.HE},
       adsurl = {https://ui.adsabs.harvard.edu/abs/2017A&A...608A..17T}
}

@ARTICLE{2017JATIS...3d4003M,
       author = {{Madsen}, Kristin K. and {Christensen}, Finn E. and {Craig}, William W. and {Forster}, Karl W. and {Grefenstette}, Brian W. and {Harrison}, Fiona A. and {Miyasaka}, Hiromasa and {Rana}, Vikram},
        title = "{Observational artifacts of Nuclear Spectroscopic Telescope Array: ghost rays and stray light}",
      journal = {JATIS},
         year = 2017,
        month = oct,
       volume = {3},
          eid = {044003},
        pages = {044003},
          doi = {10.1117/1.JATIS.3.4.044003},
       adsurl = {https://ui.adsabs.harvard.edu/abs/2017JATIS...3d4003M}
}

@ARTICLE{2023JATIS...9d8001M,
       author = {{Mukhin}, Andrey and {Krivonos}, Roman and {Vikhlinin}, Alexey and {Grefenstette}, Brian and {Madsen}, Kristin and {Wik}, Daniel},
        title = "{Wavelet-based image decomposition method for NuSTAR stray light background studies}",
      journal = {JATIS},
         year = 2023,
        month = oct,
       volume = {9},
       number = {4},
          eid = {048001},
        pages = {048001},
          doi = {10.1117/1.JATIS.9.4.048001},
archivePrefix = {arXiv},
       eprint = {2310.10516},
 primaryClass = {astro-ph.IM},
       adsurl = {https://ui.adsabs.harvard.edu/abs/2023JATIS...9d8001M}
}

@ARTICLE{Harrison2013,
       author = {{Harrison}, Fiona A. and {Craig}, William W. and {Christensen}, Finn E. and {Hailey}, Charles J. and {Zhang}, William W. and {Boggs}, Steven E. and {Stern}, Daniel and {Cook}, W. Rick and {Forster}, Karl and {Giommi}, Paolo and {Grefenstette}, Brian W. and {Kim}, Yunjin and {Kitaguchi}, Takao and {Koglin}, Jason E. and {Madsen}, Kristin K. and {Mao}, Peter H. and {Miyasaka}, Hiromasa and {Mori}, Kaya and {Perri}, Matteo and {Pivovaroff}, Michael J. and {Puccetti}, Simonetta and {Rana}, Vikram R. and {Westergaard}, Niels J. and {Willis}, Jason and {Zoglauer}, Andreas and {An}, Hongjun and {Bachetti}, Matteo and {Barri{\`e}re}, Nicolas M. and {Bellm}, Eric C. and {Bhalerao}, Varun and {Brejnholt}, Nicolai F. and {Fuerst}, Felix and {Liebe}, Carl C. and {Markwardt}, Craig B. and {Nynka}, Melania and {Vogel}, Julia K. and {Walton}, Dominic J. and {Wik}, Daniel R. and {Alexander}, David M. and {Cominsky}, Lynn R. and {Hornschemeier}, Ann E. and {Hornstrup}, Allan and {Kaspi}, Victoria M. and {Madejski}, Greg M. and {Matt}, Giorgio and {Molendi}, Silvano and {Smith}, David M. and {Tomsick}, John A. and {Ajello}, Marco and {Ballantyne}, David R. and {Balokovi{\'c}}, Mislav and {Barret}, Didier and {Bauer}, Franz E. and {Blandford}, Roger D. and {Brandt}, W. Niel and {Brenneman}, Laura W. and {Chiang}, James and {Chakrabarty}, Deepto and {Chenevez}, Jerome and {Comastri}, Andrea and {Dufour}, Francois and {Elvis}, Martin and {Fabian}, Andrew C. and {Farrah}, Duncan and {Fryer}, Chris L. and {Gotthelf}, Eric V. and {Grindlay}, Jonathan E. and {Helfand}, David J. and {Krivonos}, Roman and {Meier}, David L. and {Miller}, Jon M. and {Natalucci}, Lorenzo and {Ogle}, Patrick and {Ofek}, Eran O. and {Ptak}, Andrew and {Reynolds}, Stephen P. and {Rigby}, Jane R. and {Tagliaferri}, Gianpiero and {Thorsett}, Stephen E. and {Treister}, Ezequiel and {Urry}, C. Megan},
        title = "{The Nuclear Spectroscopic Telescope Array (NuSTAR) High-energy X-Ray Mission}",
      journal = {\apj},
         year = 2013,
        month = jun,
       volume = {770},
       number = {2},
          eid = {103},
        pages = {103},
          doi = {10.1088/0004-637X/770/2/103},
archivePrefix = {arXiv},
       eprint = {1301.7307},
 primaryClass = {astro-ph.IM},
       adsurl = {https://ui.adsabs.harvard.edu/abs/2013ApJ...770..103H}
}

@ARTICLE{Illarionov1975,
       author = {{Illarionov}, A.~F. and {Sunyaev}, R.~A.},
        title = "{Why the Number of Galactic X-ray Stars Is so Small?}",
      journal = {\aap},
         year = 1975,
        month = feb,
       volume = {39},
        pages = {185},
       adsurl = {https://ui.adsabs.harvard.edu/abs/1975A&A....39..185I}
}

@ARTICLE{Campana2002,
       author = {{Campana}, S. and {Stella}, L. and {Israel}, G.~L. and {Moretti}, A. and {Parmar}, A.~N. and {Orlandini}, M.},
        title = "{The Quiescent X-Ray Emission of Three Transient X-Ray Pulsars}",
      journal = {\apj},
         year = 2002,
        month = nov,
       volume = {580},
       number = {1},
        pages = {389-393},
          doi = {10.1086/343074},
archivePrefix = {arXiv},
       eprint = {astro-ph/0207422},
 primaryClass = {astro-ph},
       adsurl = {https://ui.adsabs.harvard.edu/abs/2002ApJ...580..389C}
}

@INPROCEEDINGS{Mushtukov_Tsygankov_2024_Review,
       author = {{Mushtukov}, Alexander and {Tsygankov}, Sergey},
        title = "{Accreting strongly magnetised neutron stars: X-ray Pulsars}",
    booktitle = {Handbook of X-ray and Gamma-ray Astrophysics}, 
publisher = {Springer},
 address = {Singapore}, 
    editor = {{Bambi}, C. and {Santangelo}, A.},
         year = 2024,
        month = apr,
        pages = {4105-4176},
          doi = {10.1007/978-981-19-6960-7_104},
archivePrefix = {arXiv},
       eprint = {2204.14185},
 primaryClass = {astro-ph.HE},
       adsurl = {https://ui.adsabs.harvard.edu/abs/2022arXiv220414185M},
      note = {\href{https://arxiv.org/abs/2204.14185}{arXiv:2204.14185}}
}

@ARTICLE{Ghosh1978,
       author = {{Ghosh}, P. and {Lamb}, F.~K.},
        title = "{Disk accretion by magnetic neutron stars.}",
      journal = {\apjl},
         year = 1978,
        month = jul,
       volume = {223},
        pages = {L83-L87},
          doi = {10.1086/182734},
       adsurl = {https://ui.adsabs.harvard.edu/abs/1978ApJ...223L..83G}
}

@ARTICLE{Kaastra2016,
       author = {{Kaastra}, J.~S. and {Bleeker}, J.~A.~M.},
        title = "{Optimal binning of X-ray spectra and response matrix design}",
      journal = {\aap},
         year = 2016,
        month = mar,
       volume = {587},
          eid = {A151},
        pages = {A151},
          doi = {10.1051/0004-6361/201527395},
archivePrefix = {arXiv},
       eprint = {1601.05309},
 primaryClass = {astro-ph.IM},
       adsurl = {https://ui.adsabs.harvard.edu/abs/2016A&A...587A.151K}
}
\bibliographystyle{aa}

\begin{appendix}
\section{Treatment of stray-light contamination}
\label{app:straylight}

Stray-light contamination \citep{2017JATIS...3d4003M} from a nearby bright X-ray source, caused by its photons entering the detector aperture without reflection from the optics, was present in all three
\nustar\ observations. In NuObs1 and NuObs2, the source flux remained
significantly higher than the background level over the energy ranges
used in the spectral analysis. Therefore, the standard data-reduction
procedure was sufficient for these observations.

\begin{figure}[h]
\centering
\includegraphics[width=1.0\linewidth]{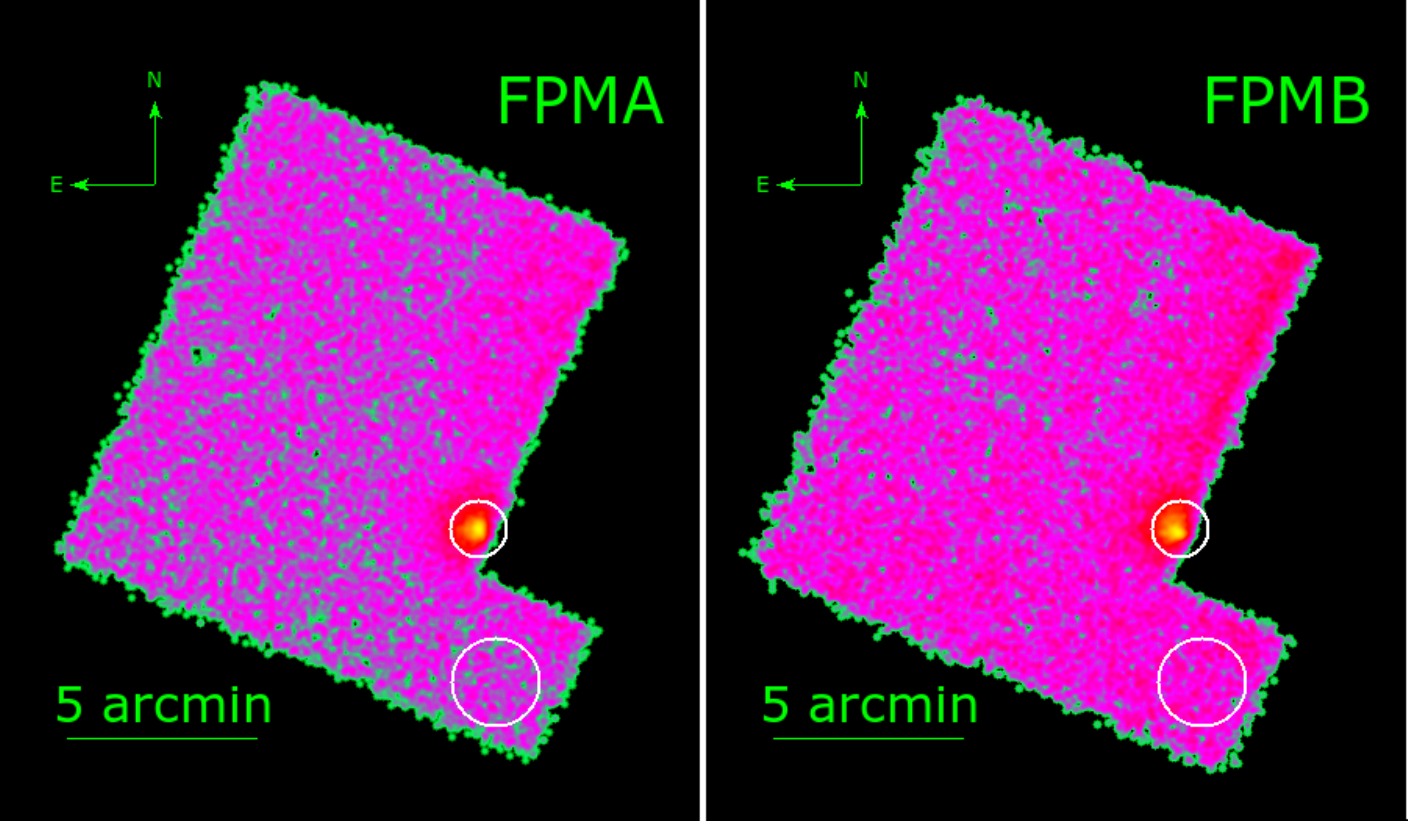}
\caption{
Smoothed, cleaned detector images from the \textit{NuSTAR} FPMA (left) and
FPMB (right) modules obtained during the faint observation
NuObs3. The images are shown after filtering detector regions
affected by stray-light contamination using custom bad-pixel
maps. The source and background extraction regions are indicated
by white circles. 
}
\label{fig:detect}
\end{figure}

In NuObs3, the stray-light contribution produced a noticeable hard
excess above $\sim30$~keV. To minimize its effect, we applied
additional filtering at the event-processing stage by supplying custom
bad-pixel maps to the \texttt{nupipeline} task through the
\texttt{fpma\_userbpfile} and \texttt{fpmb\_userbpfile} parameters, which allow \texttt{nupipeline} to apply necessary point-spread-function corrections for spectra extraction.
The custom bad-pixel maps were constructed using the \texttt{nuwavdet} package\footnote{\url{https://heagit.cosmos.ru/nustar/nuwavdet}} \citep{2023JATIS...9d8001M}, which was used to convert detector-plane masks into RAW-coordinate bad-pixel maps compatible with \texttt{nupipeline}.
The resulting cleaned FPMA and FPMB detector images are shown in
Fig.~\ref{fig:detect}.

Despite this filtering, a residual stray-light contribution remained
at high energies, where the instrumental background also increases.
We therefore restricted the NuObs3 spectral analysis to the
3--40~keV energy range. Within this interval, the source emission
remains above the background level.

\end{appendix}
\end{document}